\documentclass{article}
\usepackage{spconf,amsmath,amssymb,graphicx,booktabs}
\usepackage[protrusion=false]{microtype}
\usepackage[hidelinks]{hyperref}
\makeatletter

\long\def\@makefntext#1{\parindent 1em\noindent
  \hb@xt@1.4em{\hss\@makefnmark\hspace{0.25em}}#1}
\long\def\@makecaption#1#2{\vskip 3pt
  \setbox\@tempboxa\hbox{#1. #2}%
  \ifdim\wd\@tempboxa>\hsize #1. #2\par
  \else\hbox to\hsize{\hfil\box\@tempboxa\hfil}\fi}
\makeatother
\hypersetup{pdftitle={WhisperVC-AV: Audio-Visual Content Restoration for Noise-Robust Whisper-to-Normal Voice Conversion}, pdfauthor={Ziyue Yin, Dong Liu, Ming Li}, pdfkeywords={Whisper-to-normal voice conversion, audio-visual speech processing, content restoration, noise robustness}}
\title{WhisperVC-AV: Audio-Visual Content Restoration for\\
Noise-Robust Whisper-to-Normal Voice Conversion}
\name{Ziyue Yin$^{1,2}$, Dong Liu$^{3,4}$, Ming Li$^{4}$\sthanks{Corresponding author: Ming Li (email: mingli369@cuhk.edu.cn).}}
\address{$^1$Digital Innovation Research Center, Duke Kunshan University, Kunshan, China\\
$^2$Department of Computer Science, Johns Hopkins University, Baltimore, MD, USA\\
$^3$School of Computer Science, Wuhan University, Wuhan, China\\
$^4$School of Artificial Intelligence, The Chinese University of Hong Kong, Shenzhen, China}

\begin{document}
\ninept
\raggedbottom
\setcounter{footnote}{1}
\maketitle
\begin{abstract}
Environmental noise obscures the content cues needed for whisper-to-normal
conversion. We propose WhisperVC-AV, which restores acoustic content features
while leaving the original WhisperVC conversion module unchanged. Its context-guided
restoration module combines temporal acoustic context with synchronized lip
features through attention and gated residual correction.
Experiments on AISHELL6-Whisper show lower character error rates (CERs)
than WhisperVC across three ASR systems, on clean speech and under all six
signal-to-noise ratio (SNR) conditions with MUSAN noise. The largest gains
occur at 0 dB SNR, where Qwen3-ASR CER falls from 36.29\% to 28.88\%.
WhisperVC-AV also improves predicted speech quality while maintaining
speaker similarity. The CER gains extend to unseen background noise without retraining, while visual
controls support the use of utterance-specific lip cues. Audio examples are
available on our demo page.\footnote{Demo page: \href{https://larry-ziyue-yin.github.io/demo-whispervc-av/}{\nolinkurl{larry-ziyue-yin.github.io/demo-whispervc-av/}}.}\looseness=-1
\end{abstract}
\begin{keywords}
Whisper-to-normal voice conversion, audio-visual speech processing, content restoration, noise robustness
\end{keywords}

\begin{figure*}[t]
  \centering
  \includegraphics[width=0.99\linewidth,trim=0bp 14bp 0bp 29bp,clip]{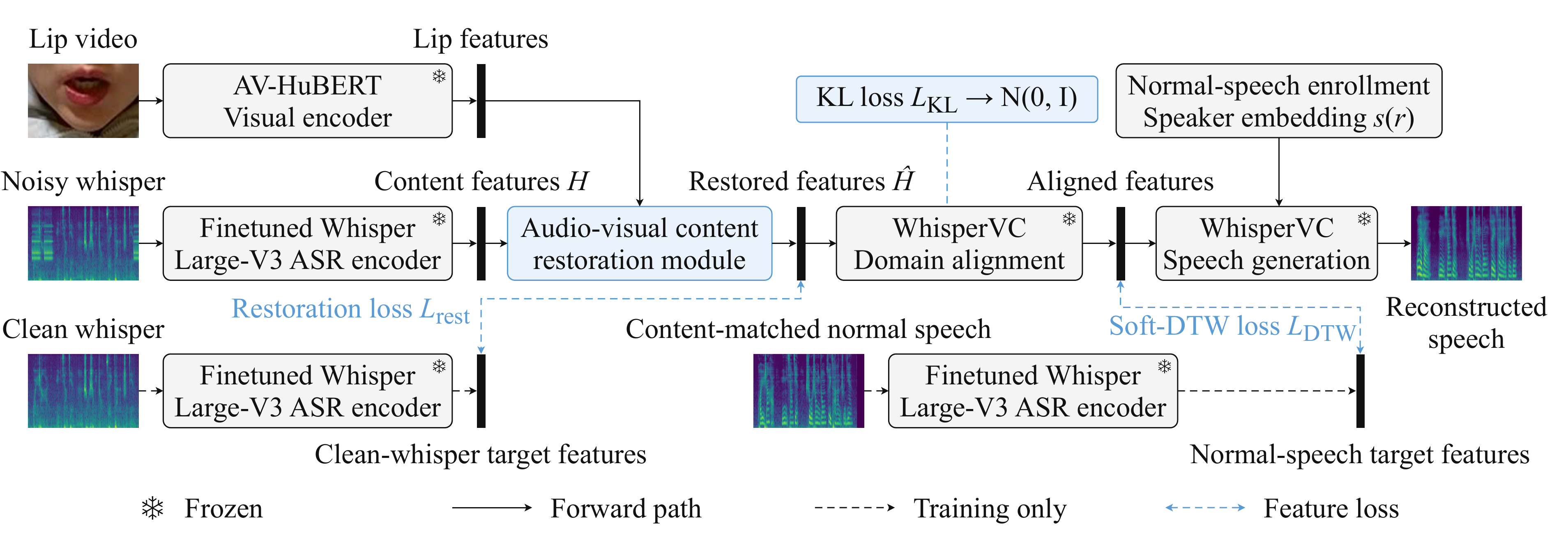}
  \par
  \caption{Overview of WhisperVC-AV. Acoustic and visual cues restore content features while the original WhisperVC conversion module remains unchanged. All speech branches share the same frozen ASR encoder. Dashed paths and feature-loss connections are used only during training.\looseness=-1}
  \label{fig:architecture}
\end{figure*}

\begin{figure*}[t]
  \centering
  \includegraphics[width=0.985\linewidth,trim=0bp 15bp 0bp 26bp,clip]{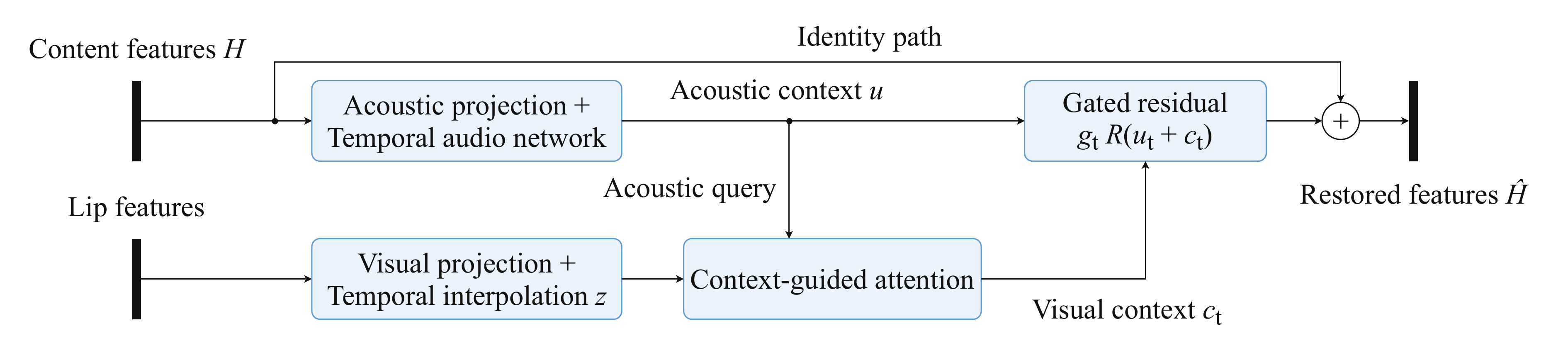}
  \par
  \caption{Context-guided content restoration. Acoustic queries attend to projected lip features $z$ within $\mathcal N(t)$. Acoustic context $u_t$ and visual context $c_t$ jointly determine the gated residual, which the identity path adds to content features $H$ to obtain restored features $\hat H$.}
  \label{fig:restoration-module}
\end{figure*}

\section{Introduction}
\label{sec:intro}
Whisper-to-normal (W2N) conversion converts whispered speech into normal
speech while preserving linguistic content and speaker identity. A speaker
may whisper to avoid disturbing others, yet ventilation, household activity,
or passing vehicles can remain audible. Whispered speech
lacks periodic vocal-fold excitation~\cite{perrotin2020glottal}, and noise
further obscures its acoustic content cues. Therefore, W2N conversion must
address both cross-phonation differences and noisy input.\looseness=-1

Earlier W2N methods learn adversarial mappings~\cite{gao2023agan}, use
variational autoencoders~\cite{seki2023acvae}, or model source--filter
synthesis~\cite{perrotin2020glottal}. More recently, WESPER and DistillW2N
leverage self-supervised speech
representations~\cite{rekimoto2023wesper,tan2025distillw2n}, while
\mbox{WhispEar} expands training data with pseudo-parallel whispered
speech~\cite{fang2026whispear}. WhisperVC instead separates cross-phonation
alignment from speech generation~\cite{liu2026whispervc}. When noise corrupts
its input features, the converter receives less reliable content information.
This motivates us to restore the input representation while retaining the
learned conversion mapping.\looseness=-1

Synchronized lip motion is a useful source of complementary information
because additive acoustic noise does not directly corrupt the video.
This complementarity has been exploited in audio-visual speech
enhancement~\cite{ephrat2018looking}, resynthesis~\cite{yang2022avcodec}, and
representation learning~\cite{shi2022avhubert,shi2022robust}.
For speech recognition, V-CAFE uses visual-context attention and
mask-plus-residual acoustic enhancement~\cite{hong2022vcafe}.
Similarly, gated visual attention incorporates lip information into
\mbox{Whisper}~\cite{li2026aishell6}. These approaches use visual evidence to
strengthen acoustic representations for recognition.

Lip information has also been integrated directly into W2N conversion:
W2N-AVSC trains a conditional VAE with a video-conditioned prior and sums
acoustic and visual latent representations for
decoding~\cite{seki2023w2navsc}.
It evaluates audiovisual conversion with both clean and noisy whispered input.
Rather than jointly learning audiovisual conversion, we use lip information to
restore the acoustic representation supplied to a pretrained converter,
preserving its learned conversion mapping. Our contributions are:\looseness=-1

\begin{list}{\textbullet}{\setlength{\leftmargin}{1em}
\setlength{\labelwidth}{0.6em}\setlength{\labelsep}{0.4em}
\setlength{\topsep}{2pt}\setlength{\itemsep}{2pt}
\setlength{\parsep}{0pt}\setlength{\partopsep}{0pt}}
\item We formulate \textbf{noise-robust W2N conversion} as \textbf{input-side content
restoration}, reusing a pretrained converter without updating its conversion
backbone.
\item \begin{samepage}We develop a \textbf{context-guided audio-visual restoration module}
that uses acoustic context to select nearby lip information and predicts a
gated residual correction.\looseness=-1\par\end{samepage}
\end{list}

By restoring content at the converter's input, we separate noise robustness
from the learned whisper-to-normal mapping. This design enables us to reuse
WhisperVC's conversion capability while exploiting synchronized lip cues
when acoustic evidence is degraded.

\section{Method}
\label{sec:method}
\subsection{Overview}
Fig.~\ref{fig:architecture} shows how we separate input restoration from conversion.
For noisy whispered speech $\tilde{x}$, a frozen Whisper Large-V3 encoder
finetuned on AISHELL6-Whisper~\cite{li2026aishell6} extracts acoustic content
features $H$. Each utterance is instance-normalized over time per feature
channel, with frozen affine parameters inherited from WhisperVC. Clean
training targets are normalized independently. Our module combines $H$ with
synchronized lip features to produce $\hat H$ in the same acoustic space.
WhisperVC's alignment network maps $\hat H$ to normal-speech content
features; its generator and vocoder reconstruct speech $\hat y$, conditioned
on embedding $s(r)$ from same-speaker normal-speech enrollment $r$~\cite{liu2026whispervc}.
Only the restoration module is trained; the feature encoders and conversion
backbone remain frozen.\looseness=-1

\subsection{Context-guided audio-visual restoration}
In Fig.~\ref{fig:restoration-module}, a projection maps 1280-dimensional $H$
to 256 dimensions. Three residual temporal convolutional layers, with
kernel size 5 and dilations 1, 2, and 4, form acoustic context $u$.
Frozen noise-robust AV-HuBERT~\cite{shi2022robust}, with audio disabled,
extracts 1024-dimensional lip features from 25-fps video. Layer normalization
and a learned projection map them to 256 dimensions; linear interpolation
aligns them with the 50-Hz acoustic sequence, yielding $z$.
Single-head scaled dot-product attention takes $u$ as queries and $z$ as
both keys and values. Each query
attends to $\mathcal N(t)=\{t-4,\ldots,t+4\}$, spanning approximately
$\pm80$ ms; invalid boundary positions are masked.
Acoustic context $u_t$ and visual context $c_t$ jointly determine a scalar
sigmoid gate $g_t$ and the residual correction,
\begin{equation}
 \begin{aligned}
 c_t &= \mathrm{Attn}\bigl(u_t,\{z_j:j\in\mathcal{N}(t)\}\bigr),\\
 g_t &= \sigma\bigl(W_g[u_t;c_t]+b_g\bigr),\quad
 \hat{H}_t = H_t+g_tR(u_t+c_t).
 \end{aligned}
 \label{eq:restoration}
\end{equation}
Here $[u_t;c_t]$ denotes concatenation, and $R$ projects back to the input
feature space. Zero-initializing $R$ preserves the identity output.
\looseness=-1

\begin{table*}[t]
\centering
\caption{Character error rates (CER, \%) on clean speech and under six SNR conditions with MUSAN noise. The cascades enhance waveforms before WhisperVC. WhisperVC-A is the separately trained audio-only ablation. Noisy avg. averages the six noisy conditions; lower is better, with each condition's best result in bold.}
\label{tab:main}
\vspace{3pt}
\small
\renewcommand{\arraystretch}{0.92}
\setlength{\tabcolsep}{2.2pt}
\begin{tabular*}{\linewidth}{@{\extracolsep{\fill}}llrrrrrrrr@{}}
\toprule
ASR & System & Clean & 0 dB & 5 dB & 10 dB & 15 dB & 20 dB & 25 dB & Noisy avg.\\
\midrule
Qwen3-ASR & WhisperVC~\cite{liu2026whispervc} & 11.47 & 36.29 & 26.23 & 18.91 & 15.40 & 13.43 & 12.34 & 20.43\\
 & MP-SENet-DNS + WhisperVC & 24.42 & 43.13 & 34.31 & 26.97 & 22.57 & 19.72 & 17.90 & 27.43\\
 & SEMamba + WhisperVC & 25.34 & 61.63 & 52.04 & 43.18 & 35.67 & 31.63 & 28.71 & 42.14\\
 & Direct AV adaptation & 10.40 & 29.99 & 21.51 & 16.18 & 13.20 & 11.65 & 11.05 & 17.26\\
 & WhisperVC-A & 10.11 & 30.92 & 21.76 & 16.08 & 13.27 & 11.66 & 10.87 & 17.43\\
 & \textbf{WhisperVC-AV (ours)} & \textbf{9.86} & \textbf{28.88} & \textbf{20.60} & \textbf{15.31} & \textbf{12.77} & \textbf{11.36} & \textbf{10.57} & \textbf{16.58}\\
\midrule
Finetuned Whisper & WhisperVC~\cite{liu2026whispervc} & 13.68 & 52.97 & 36.36 & 23.92 & 17.70 & 15.99 & 14.55 & 26.92\\
(Large-V3) in~\cite{li2026aishell6} & MP-SENet-DNS + WhisperVC & 33.18 & 65.47 & 47.08 & 34.03 & 27.76 & 24.44 & 22.76 & 36.92\\
 & SEMamba + WhisperVC & 30.11 & 92.71 & 80.30 & 57.61 & 44.15 & 36.79 & 35.86 & 57.90\\
 & Direct AV adaptation & 12.24 & 41.36 & 26.28 & 18.31 & 15.18 & 14.17 & 12.85 & 21.36\\
 & WhisperVC-A & 11.46 & 33.46 & 23.15 & 17.36 & 14.52 & 12.94 & 12.28 & 18.95\\
 & \textbf{WhisperVC-AV (ours)} & \textbf{11.27} & \textbf{30.43} & \textbf{21.45} & \textbf{16.83} & \textbf{13.97} & \textbf{12.52} & \textbf{11.90} & \textbf{17.85}\\
\midrule
Paraformer & WhisperVC~\cite{liu2026whispervc} & 14.01 & 39.98 & 29.81 & 22.27 & 18.63 & 16.44 & 15.20 & 23.72\\
 & MP-SENet-DNS + WhisperVC & 28.13 & 46.68 & 38.16 & 31.04 & 26.25 & 23.22 & 21.27 & 31.10\\
 & SEMamba + WhisperVC & 29.57 & 64.10 & 55.21 & 47.25 & 39.71 & 35.72 & 32.81 & 45.80\\
 & Direct AV adaptation & 12.96 & 34.00 & 25.56 & 19.50 & 16.30 & 14.70 & 13.70 & 20.63\\
 & WhisperVC-A & 12.68 & 35.14 & 25.60 & 19.61 & 16.29 & 14.41 & 13.57 & 20.77\\
 & \textbf{WhisperVC-AV (ours)} & \textbf{12.33} & \textbf{33.20} & \textbf{24.73} & \textbf{19.10} & \textbf{15.80} & \textbf{14.10} & \textbf{13.23} & \textbf{20.03}\\
\bottomrule
\end{tabular*}
\end{table*}

\subsection{Training and inference}
Training pairs noisy whispered speech with its clean counterpart $x$.
Noisy training inputs undergo waveform time masking with probability 0.5
before feature extraction; clean inputs are unmasked. We zero randomly
placed spans of 0.2--0.6 s until 20\% of waveform samples are masked,
merging overlaps and truncating spans to the remaining budget.
One masked variant is cached per noisy input. Feature-frame masks sample
the waveform mask at 20-ms intervals to match the acoustic sequence.
Restoration loss
$\mathcal L_{\mathrm{rest}}$ weights mean Smooth-L1 errors on masked and
unmasked frames by 0.75 and 0.25, respectively, against clean features from
the shared frozen encoder and normalization. Without masked frames, it uses
the overall mean. Masks guide the loss but are not module inputs.
For downstream compatibility, we retain WhisperVC's alignment
and latent objectives~\cite{liu2026whispervc}. Soft-DTW~\cite{cuturi2017softdtw}
compares aligned features with content-matched normal-speech targets without
assuming identical timing. KL regularization constrains the whisper-path
posterior toward $\mathcal N(0,I)$. Figure~\ref{fig:architecture} shows the
loss paths for the objective
\begin{equation}
\mathcal{L}=\mathcal{L}_{\mathrm{DTW}}
 +\lambda_{\mathrm{KL}}\mathcal{L}_{\mathrm{KL}}
 +\lambda_{\mathrm{rest}}\mathcal{L}_{\mathrm{rest}}.
\label{eq:objective}
\end{equation}

We set $\lambda_{\mathrm{KL}}=0.1$ to keep KL regularization modest and
$\lambda_{\mathrm{rest}}=50$ to give the smaller, frame-averaged restoration
loss an effective contribution alongside Soft-DTW. Although the alignment
network is frozen, it remains differentiable: gradients from its Soft-DTW
and KL losses pass through $\hat H$ to update the restoration module, while
its own weights remain fixed. At inference, only noisy speech, lip video, and speaker
enrollment are needed (Fig.~\ref{fig:architecture}).\looseness=-1

\section{Experimental results}
\label{sec:experiments}
\subsection{Experimental setup}
\label{sec:setup}
We retain AISHELL6-Whisper's original partition boundaries~\cite{li2026aishell6}
and use its video-covered subsets: 9,494 training utterances from 81 speakers,
2,235 validation utterances from 19 speakers, and the same 2,248 test
utterances from 20 speakers, under clean input and each noisy condition.
The utterance lists are available on our demo page.

All WhisperVC baseline metrics are recomputed on this subset; the original
paper used a broader test set and Whisper-large-v3-turbo for
CER~\cite{liu2026whispervc}.
FreeSound and SoundBible recordings from MUSAN's
\emph{noise} subset~\cite{snyder2015musan} are mixed at 0, 5, 10, 15, 20,
and 25 dB SNR, including environmental sounds such as crowd noise,
dial tones, and squeaking doors. Music and speech subsets are excluded,
and noise files are disjoint across partitions. Noise files and starting
offsets are sampled separately for each SNR. Within each condition, all
evaluated pipelines start from the same noisy recording of each test utterance.
We also evaluate clean input.\looseness=-1

Content-matched normal recordings from the same speaker provide training
targets without frame-level alignment. The enrollment speech, however, uses
a different sentence spoken normally by the same speaker, during both
training and inference. For evaluation, each test utterance uses the same
enrollment recording across systems and SNR conditions. Targets and
enrollment remain clean.

\subsection{Evaluation metrics}
Character error rate (CER) measures intelligibility and content
preservation. Three ASR systems are used to assess the consistency of CER
findings: Qwen3-ASR~\cite{shi2026qwen3asr}, Finetuned Whisper (Large-V3)
in~\cite{li2026aishell6}, and Paraformer~\cite{gao2022paraformer}.
For each condition, we sum character substitutions, deletions, and insertions
over all test utterances and divide by the total number of reference characters.\looseness=-1

Predicted output quality is assessed with DNSMOS P.835~\cite{reddy2022dnsmos},
UTMOS~\cite{saeki2022utmos}, WVMOS~\cite{andreev2023hifipp}, and
NISQA~\cite{mittag2021nisqa}. Resemblyzer SECS,
WeSpeaker~\cite{wang2024wespeaker}, and
WavLM~\cite{chen2022wavlm} measure similarity to content-matched normal
recordings; SpeechBERTScore~\cite{saeki2024speechbert} evaluates consistency
with the same references.

\begin{table}[t]
\centering
\caption{Mean CER reductions of WhisperVC-AV over six SNR conditions with MUSAN noise (percentage points), with pointwise 95\% confidence intervals (CIs). All nine two-sided paired tests yield raw $p=0.000005$ (199,999 label swaps; plus-one correction) and Holm-adjusted $p=0.000045$.}
\label{tab:significance}
\vspace{3pt}
\small
\renewcommand{\arraystretch}{0.92}
\setlength{\tabcolsep}{2pt}
\begin{tabular*}{\linewidth}{@{\extracolsep{\fill}}llr@{}}
\toprule
ASR & Compared with & Reduction [95\% CI]\\
\midrule
Qwen3-ASR & WhisperVC & 3.85 [3.57, 4.13]\\
 & Direct AV adaptation & 0.68 [0.48, 0.88]\\
 & WhisperVC-A & 0.85 [0.73, 0.97]\\
\midrule
Finetuned Whisper & WhisperVC & 9.07 [7.76, 10.46]\\
(Large-V3) in~\cite{li2026aishell6} & Direct AV adaptation & 3.51 [2.64, 4.47]\\
 & WhisperVC-A & 1.10 [0.69, 1.54]\\
\midrule
Paraformer & WhisperVC & 3.69 [3.44, 3.95]\\
 & Direct AV adaptation & 0.60 [0.40, 0.80]\\
 & WhisperVC-A & 0.74 [0.64, 0.85]\\
\bottomrule
\end{tabular*}
\end{table}

\begin{table*}[t]
  \begingroup
\caption{Predicted speech quality, speaker similarity, and reference consistency for clean input and the average over six SNR conditions with MUSAN noise. Higher values are better; bold indicates the best rounded score within each condition.}
\label{tab:quality}
\centering
\vspace{3pt}
\small
\renewcommand{\arraystretch}{0.92}
\setlength{\tabcolsep}{3pt}
\begin{tabular*}{\linewidth}{@{\extracolsep{\fill}}llrrrrrrrrrrr@{}}
\toprule
 & & \multicolumn{4}{c}{DNSMOS $\uparrow$} & \multicolumn{3}{c}{MOS predictors $\uparrow$} & \multicolumn{4}{c}{Speaker/reference similarity $\uparrow$}\\
\cmidrule(lr){3-6}\cmidrule(lr){7-9}\cmidrule(l){10-13}
System & Set & OVRL & SIG & BAK & P808 & UTMOS & WVMOS & NISQA & SECS & WeSpk & WavLM & SBERT\\
\midrule
WhisperVC~\cite{liu2026whispervc} & Clean & 3.07 & 3.51 & 3.72 & 3.58 & 2.83 & 3.29 & 3.13 & 0.83 & 0.68 & 0.94 & 0.78\\
 & Noisy avg. & 3.05 & 3.48 & 3.71 & 3.54 & 2.76 & 3.22 & 3.11 & 0.82 & 0.67 & \textbf{0.94} & 0.77\\
Direct AV adaptation & Clean & 3.09 & 3.52 & 3.73 & 3.59 & 2.85 & \textbf{3.30} & 3.15 & \textbf{0.84} & \textbf{0.69} & \textbf{0.95} & \textbf{0.79}\\
 & Noisy avg. & 3.08 & 3.52 & 3.73 & 3.58 & 2.82 & \textbf{3.28} & 3.14 & \textbf{0.83} & 0.68 & \textbf{0.94} & \textbf{0.78}\\
WhisperVC-A & Clean & \textbf{3.17} & \textbf{3.55} & \textbf{3.84} & \textbf{3.67} & \textbf{2.90} & 3.28 & \textbf{3.18} & \textbf{0.84} & \textbf{0.69} & \textbf{0.95} & \textbf{0.79}\\
 & Noisy avg. & \textbf{3.16} & \textbf{3.55} & \textbf{3.84} & \textbf{3.67} & \textbf{2.89} & \textbf{3.28} & \textbf{3.18} & \textbf{0.83} & \textbf{0.69} & \textbf{0.94} & \textbf{0.78}\\
\textbf{WhisperVC-AV (ours)} & Clean & 3.16 & \textbf{3.55} & 3.83 & \textbf{3.67} & \textbf{2.90} & 3.28 & \textbf{3.18} & \textbf{0.84} & \textbf{0.69} & \textbf{0.95} & \textbf{0.79}\\
 & Noisy avg. & \textbf{3.16} & \textbf{3.55} & \textbf{3.84} & \textbf{3.67} & \textbf{2.89} & 3.27 & \textbf{3.18} & \textbf{0.83} & \textbf{0.69} & \textbf{0.94} & \textbf{0.78}\\
\bottomrule
\end{tabular*}
\par\endgroup

  \par\vspace{3pt}
  \noindent
  \begin{minipage}[t]{\dimexpr(\textwidth-\columnsep)/2\relax}
    \vspace{0pt}
\begingroup
\centering
\caption{W2N CER (\%) with unseen background noise from DEMAND at three SNR levels. Lower is better.\strut}
\label{tab:demand}
\vspace{3pt}
\small
\renewcommand{\arraystretch}{0.92}
\setlength{\tabcolsep}{0.8pt}
\begin{tabular*}{\linewidth}{@{\extracolsep{\fill}}lrrrrrr@{}}
\toprule
& \multicolumn{3}{c}{Qwen3-ASR} & \multicolumn{3}{c}{\shortstack{Finetuned Whisper\\(Large-V3) in~\cite{li2026aishell6}}}\\
\cmidrule(lr){2-4}\cmidrule(l){5-7}
System & 0 dB & 5 dB & 10 dB & 0 dB & 5 dB & 10 dB\\
\midrule
WhisperVC~\cite{liu2026whispervc} & 24.72 & 17.26 & 13.78 & 30.80 & 23.61 & 18.24\\
WhisperVC-A & 20.80 & 14.75 & 12.06 & 21.78 & 15.88 & 13.22\\
\textbf{WhisperVC-AV (ours)} & \textbf{19.71} & \textbf{14.22} & \textbf{11.90} & \textbf{20.43} & \textbf{15.39} & \textbf{13.17}\\
\bottomrule
\end{tabular*}
\par\endgroup

  \end{minipage}\hfill
  \begin{minipage}[t]{\dimexpr(\textwidth-\columnsep)/2\relax}
    \vspace{0pt}
    \begingroup
\centering
\caption{Qwen3-ASR CER (\%) for visual controls on clean speech and under six SNR conditions with MUSAN noise. Lower is better.\strut}
\label{tab:controls}
\vspace{3pt}
\small
\renewcommand{\arraystretch}{0.92}
\setlength{\tabcolsep}{2.5pt}
\begin{tabular*}{\linewidth}{@{\extracolsep{\fill}}lrrrrrrr@{}}
\toprule
\multicolumn{8}{c}{\textbf{WhisperVC-AV (ours)}}\\
Video & Clean & 0 dB & 5 dB & 10 dB & 15 dB & 20 dB & 25 dB\\
\midrule
Correct & \textbf{9.86} & \textbf{28.88} & \textbf{20.60} & \textbf{15.31} & \textbf{12.77} & \textbf{11.36} & \textbf{10.57}\\
Disabled & 10.22 & 31.12 & 22.14 & 16.27 & 13.37 & 11.54 & 10.69\\
Mismatch & 10.15 & 31.78 & 22.63 & 16.70 & 13.57 & 11.78 & 10.81\\
Shift $-10$ & 10.29 & 31.34 & 22.25 & 16.50 & 13.59 & 11.52 & 10.92\\
Shift $+10$ & 10.25 & 31.59 & 22.06 & 16.33 & 13.44 & 11.70 & 10.89\\
\bottomrule
\end{tabular*}
\par\endgroup

  \end{minipage}
\end{table*}

\subsection{W2N results and ablation analysis}
We evaluate WhisperVC-AV against original WhisperVC~\cite{liu2026whispervc}
and examine waveform enhancement, direct visual adaptation, and the
contribution of the visual branch (Table~\ref{tab:main}). For the last
comparison, we separately train an audio-only ablation, denoted
WhisperVC-A. It removes the visual projection and attention paths in
Fig.~\ref{fig:restoration-module}, so that the gated residual depends solely
on acoustic context $u_t$. The acoustic network, supervision, and training
schedule are retained.\looseness=-1

The enhancement cascades apply magnitude--phase
MP-SENet-DNS~\cite{lu2025mpsenet} or Mamba-based
SEMamba~\cite{chao2024semamba} before \mbox{WhisperVC}, using official
pretrained weights without task-specific fine-tuning. Both yield higher CER
than WhisperVC on clean speech and under all six SNR conditions.
MP-SENet-DNS, the stronger cascade, increases noisy-average Qwen3-ASR CER
from 20.43\% to 27.43\%, whereas WhisperVC-AV reduces it to 16.58\%.
For these off-the-shelf enhancement cascades, waveform preprocessing does
not improve W2N content preservation; task-specific feature restoration
produces lower CER across the tested conditions.

W2N-AVSC's trained model and audiovisual test data were unavailable to us
for matched evaluation. Motivated by its audiovisual fusion
strategy~\cite{seki2023w2navsc}, we implement Direct AV adaptation within
WhisperVC.
This baseline trains gated visual fusion and the
whispered-speech alignment branch with Soft-DTW and KL, using the same
training subset and visual features; other modules remain frozen.
It improves on WhisperVC, but WhisperVC-AV achieves lower CER on clean
speech and at every tested SNR under all three scorers. This system-level
comparison supports the effectiveness of restoring acoustic content while
retaining the pretrained conversion module.

The WhisperVC-A ablation tests whether visual information improves on
audio-only restoration. Adding the visual branch lowers noisy-average
Qwen3-ASR CER from 17.43\% to 16.58\% and clean CER from 10.11\% to 9.86\%.
Its largest gain occurs at 0 dB SNR, from 30.92\% to 28.88\%, compared with
10.87\% to 10.57\% at 25 dB SNR. Finetuned Whisper and Paraformer show
the same A-to-AV improvement in every condition. Thus, synchronized lip
cues contribute beyond the acoustic context, especially when noise makes
the speech signal least reliable.

We assess six-SNR average CER gains using paired label-swap tests and
5,000 paired bootstrap draws, grouping each utterance's six conditions.
As shown in Table~\ref{tab:significance}, gains over
WhisperVC, Direct AV, and WhisperVC-A are significant under all three
scorers after Holm correction. The positive 95\% confidence intervals (CIs)
support consistent gains; the Qwen3-ASR visual gain is 0.85 percentage
points, with a CI of [0.73, 0.97]. WhisperVC-AV improves on WhisperVC-A for 20, 19, and 20
of the 20 test speakers under the respective scorers. Resampling speakers gives 95\%
CIs of [0.59, 1.12], [0.63, 1.64], and [0.53, 0.98] percentage points.
All remain positive, indicating that the visual benefit is not driven by
only a few speakers.

Table~\ref{tab:quality} shows that the CER gains are accompanied by improved
predicted speech quality. Relative to WhisperVC, WhisperVC-AV raises
noisy-average DNSMOS OVRL from 3.05 to 3.16 and UTMOS from 2.76 to 2.89.
DNSMOS, UTMOS, and NISQA also improve on clean input, while speaker/reference
similarity scores are maintained or slightly higher. WhisperVC-A and
WhisperVC-AV achieve comparable quality and similarity scores, while
WhisperVC-AV further reduces CER. Thus, visual cues improve content preservation
while retaining the quality gains of audio-only restoration.

\subsection{Zero-shot noise generalization}
\label{sec:demand}
Without retraining or adaptation, we evaluate the W2N performance of
WhisperVC, the WhisperVC-A ablation, and WhisperVC-AV on whispered speech mixed with
unseen background noise from DEMAND~\cite{thiemann2013demand}. We use kitchen
(DKITCHEN), subway-station (PSTATION), and bus (TBUS) recordings at 0, 5,
and 10 dB SNR, following the same noise-mixing protocol as MUSAN.
WhisperVC-AV outperforms WhisperVC at all three SNRs under both ASR
scorers (Table~\ref{tab:demand}). Qwen3-ASR CER falls from 24.72\% to
19.71\% at 0 dB SNR and from 13.78\% to 11.90\% at 10 dB SNR.
WhisperVC-AV also improves on the audio-only ablation at every SNR under
both scorers, reducing three-SNR mean CER by 0.59 and 0.63 percentage
points for Qwen3-ASR and Finetuned Whisper, respectively. These visual
gains extend to unseen noise and are largest at 0 dB SNR, where acoustic
evidence is most degraded.

\subsection{Visual-input controls}
We probe WhisperVC-AV's use of visual information with
parameters and acoustic input fixed. Same-speaker mismatch
preserves identity but replaces the utterance's lip movements, while shifts
of $\pm10$ native frames (approximately $\pm400$ ms), with boundary
padding, alter temporal correspondence.
Unlike the separately trained WhisperVC-A ablation, the disabled-context control
retains WhisperVC-AV's trained parameters and sets $c_t=0$ only at inference.\looseness=-1

Every control increases Qwen3-ASR CER for clean speech and all six SNR
conditions (Table~\ref{tab:controls}). At 0 dB SNR, CER rises from 28.88\%
to 31.12\% without visual context and to 31.78\% with same-speaker mismatch.
Mismatch is also worse in every other noisy condition: incorrect lip
content can be more harmful than no visual
context. Both temporal shifts also increase CER across conditions.
Together, these controls show that useful visual input must carry the
utterance's articulatory content: retaining the correct speaker alone is
insufficient. The separately trained audio-only ablation establishes the
benefit of adding vision, while these fixed-model controls establish the
importance of the supplied lip content. Their agreement supports the use
of synchronized visual evidence to complement acoustic restoration,
especially when noise obscures speech cues.\looseness=-1

\section{Conclusion}
In this paper, we propose WhisperVC-AV to restore acoustic content while
leaving the original W2N conversion module unchanged. It achieves the lowest
CER among evaluated systems, with consistent gains across three ASR scorers
on clean speech and under all six SNR conditions with MUSAN noise.
The model also improves predicted speech quality over WhisperVC while
maintaining speaker similarity. Visual cues significantly reduce
noisy-average CER over the audio-only ablation. The full system also
generalizes to unseen noise without retraining. These findings motivate low-volume and assistive communication
applications using synchronized lip video and clean enrollment.
Future work includes uncontrolled video, reverberation, noisy enrollment,
other languages, and human listening tests.\looseness=-1

\newpage
\section{Acknowledgement}
This research is funded in part by the National Natural Science Foundation of
China (62571223). The authors declare no conflicts of interest.
Large language models were used solely to polish the language of this
manuscript. All scientific content was developed and verified by the authors.\looseness=-1

\section{Compliance with Ethical Standards}
This study uses existing AISHELL6-Whisper recordings in accordance with the
data provider's authorization. No new participants were recruited or
human-subject data collected for this study. No additional ethical approval
was required for this secondary analysis.

{\small
\bibliographystyle{IEEEbib}
\let\originalbibliography\thebibliography
\renewcommand{\thebibliography}[1]{\originalbibliography{#1}\setlength{\itemsep}{1pt}}
\bibliography{refs}
}
\end{document}